\documentclass[lettersize,journal]{IEEEtran}
\usepackage{amsmath,amsfonts}
\usepackage{algorithmic}
\usepackage{algorithm}
\usepackage{array}
\usepackage{bm}
\usepackage[caption=false,font=normalsize,labelfont=sf,textfont=sf]{subfig}
\usepackage{textcomp}
\usepackage{stfloats}
\usepackage{bbding}
\usepackage{url}
\usepackage{verbatim}
\usepackage{graphicx}
\usepackage{cite}
\usepackage{multirow}
\usepackage{picinpar}
\usepackage{booktabs}

\begin{document}

\title{Turn Passive to Active: A Survey on Active Intellectual Property Protection of Deep Learning Models}

\author{Mingfu Xue, Leo Yu Zhang, Yushu Zhang, and Weiqiang Liu
\thanks{M. Xue, Y. Zhang are with the College of Computer Science and Technology, Nanjing University of Aeronautics and Astronautics, Nanjing, 211106, China (email: mingfu.xue@nuaa.edu.cn; yushu@nuaa.edu.cn).}
\thanks{L.Y. Zhang is with the School of Information and Communication Technology, Griffith University, QLD, Australia (e-mail: leocityu@gmail.com).}
\thanks{W. Liu is with the College of Electronic and Information Engineering, Nanjing University of Aeronautics and Astronautics, Nanjing, 211106, China
(e-mail: liuweiqiang@nuaa.edu.cn).}}

\maketitle

\begin{abstract}
The intellectual property protection of deep learning (DL) models has attracted increasing serious concerns. Many works on intellectual property protection for Deep Neural Networks (DNN) models have been proposed. The vast majority of existing work uses DNN watermarking to verify the ownership of the model after piracy occurs, which is referred to as passive verification. On the contrary, we focus on a new type of intellectual property protection method named active copyright protection, which refers to active authorization control and user identity management of the DNN model. As of now, there is relatively limited research in the field of active DNN copyright protection.
In this review, we attempt to clearly elaborate on the connotation, attributes, and requirements of active DNN copyright protection, provide evaluation methods and metrics for active copyright protection, review and analyze existing work on active DL model intellectual property protection, discuss potential attacks that active DL model copyright protection techniques may face, and provide challenges and future directions for active DL model intellectual property protection. This review is helpful to systematically introduce the new field of active DNN copyright protection and provide reference and foundation for subsequent work.
\end{abstract}

\begin{IEEEkeywords}
Intellectual property protection of deep learning models, Deep Neural Networks, active intellectual property protection, copyright management, active authorization control, user identity authentication and management.
\end{IEEEkeywords}

\section{Introduction}
\IEEEPARstart{S}{ince} 2017, the intellectual property (IP) protection of deep learning (DL) models has attracted more and more concerns \cite{UchidaNSS17, tai}.
Unlike traditional IP protection in the multimedia field, DL model IP protection is a new and challenging area where existing techniques cannot be directly applied \cite{tai}.
As an emerging cutting-edge research direction, Deep Neural Networks (DNN) IP protection is still in its early stages. As the deep neural networks are applied widely, studying DNN copyright protection has urgent need and important significance. In recent years, various deep learning model IP protection methods have been proposed \cite{ZhangJieAAAI, Wuhanzhou, qinchuan, qinchuan2, Quanyuuhi, XiquanGuan, UchidaNSS17, DeepSigns, Watermarking, Backdoor, CHANGTIFS, INFIP, Yinzhaoxia, ZWMtomccap, ZWMicassp, WHZisdfs, QTcompsec, WHZisdfs21}, most of which belong to the category of ``passive verification'', which involves embedding watermarks in deep learning models or extracting model signatures, and then verifying the ownership of suspicious models after piracy and infringement occur. However, this kind of passive verification method afterward cannot proactively prevent the occurrence of piracy and infringement in the first place. Recently, a small number of scholars have paid attention to proactive deep learning model IP protection methods, offering promising directions in this emerging research field.

In this review, we attempt to clearly and systematically elaborate on the connotation, attributes, and requirements of active DL intellectual property protection, provide evaluation methods and metrics for active DL copyright protection, review and analyze existing active DL model IP protection work, and discuss potential attacks that active DL model copyright protection may face. Finally, we will provide insights into the challenges and future directions for active DL model IP protection.

The contributions of this work are multi-fold:
\begin{itemize}
\item We present the first review focusing on active DL model IP protection, which fills a gap in the existing literature.
\item The connotation, attributes, requirements, objectives, evaluation suggestions and evaluation metrics of active DL model IP protection have been systematically proposed for the first time, providing valuable insights for future research.
\item We conduct a comprehensive review of existing active DL model IP protection methods, highlighting their respective advantages and disadvantages.
\item Potential attacks that active DL model IP protection may face are thoroughly discussed, enhancing the understanding of the security challenges in this domain.
\item The challenges and future directions of active DL model IP protection are presented.
\end{itemize}

This paper is organized as follows. The connotation, attributes, requirements, objectives and special metrics of active DL IP protection are proposed in Section \ref{sec_definition}. The review of existing active DL model IP protection work is presented in Section \ref{sec_review}. Section \ref{sec_attacks} discusses potential specific attacks targeted at active DL IP protection. Section \ref{sec_directions} presents the challenges and future directions of active DL model IP protection. This paper is summarized in Section \ref{sec_conclusion}.
\IEEEpubidadjcol

\section{The Connotation, Requirements, and Metrics of Active IP Protection for DL Models} \label{sec_definition}

\subsection{What is Active DL Model IP Protection?}
The work \cite{tai} provides a taxonomy of DNN IP protection methods.
From a ``type'' perspective, the taxonomy includes two categories:
(i) \textit{Passive verification}, which refers to embedding watermarks in the model and passively verifying the model's copyright when the model is suspected to be pirated. The vast majority of existing DNN IP protection works fall under this category.
(ii) \textit{Active authorization control},
which focuses on managing and controlling model usage. Only authorized users are allowed access, preventing unauthorized use. This paper specifically concentrates on this area.

\begin{figure*}[htbp]
\centering
\includegraphics[width=0.85\textwidth]{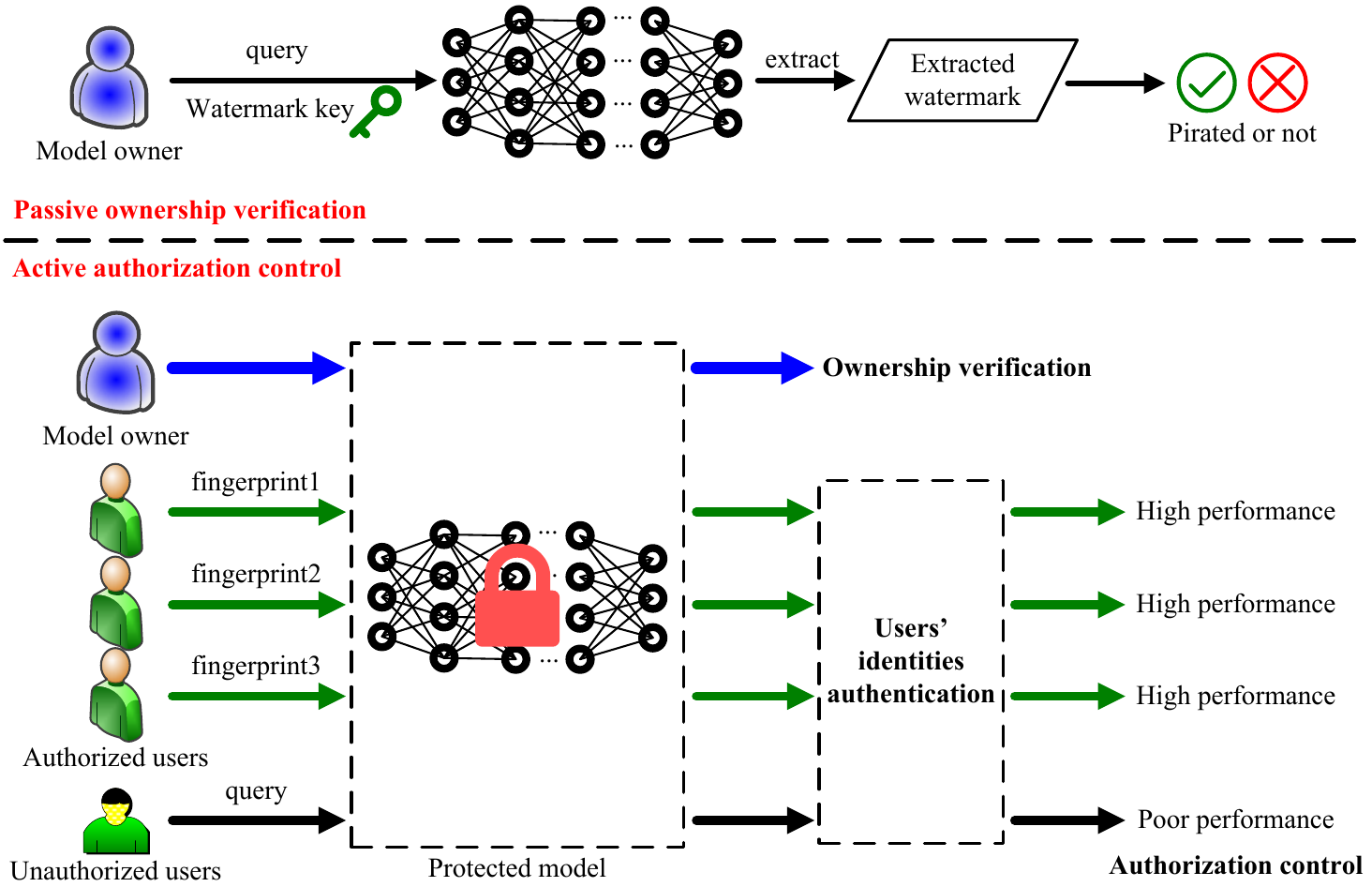}
\caption{Illustration of passive ownership verification and active authorization control.}
\label{fig_AAC}
\end{figure*}

From a ``function'' perspective, DNN copyright protection methods can be divided into \cite{tai}: (i) \textit{Copyright verification}, which means verifying the model ownership, and most existing work belongs to this category;
(ii) \textit{Copyright management}, which involves authentication and management of users' identities, as well as active authorization control of users' usage. This is a necessary function of commercial DNN and also the focus and innovation point of this paper. The illustration of passive ownership verification and active authorization control is shown in Fig.~\ref{fig_AAC}.

Numerous methods for protecting DNN copyright have emerged, yet a significant portion relies on passive verification, unable to proactively prevent piracy and infringement. While many concentrate on model ownership verification, few address user authentication and management of unique identities.
This is a necessary and key function of commercial copyright management. Besides, it is vulnerable to attacks by dishonest users, such as collusion attacks. Such vulnerabilities hinder the broad commercial use and deployment of DL models.

Given the escalating application demands for deep learning copyright protection and the need to overcome existing challenges, we deem the exploration of proactive protection methods for deep learning models imperative and vital. This review is committed to addressing the following challenges and attaining the ensuing objectives:
\begin{itemize}
\item Most of the existing work uses DNN watermarks to verify the copyright of DNN after the occurrence of piracy and infringement. Can we proactively prevent the occurrence of piracy and infringement in advance --- active authorization control?
\item How to distinguish different DNN users in order to meet the needs of commercial DNN copyright protection --- user identity authentication and management (i.e. copyright management)?
\item The various attacks considered in existing work are mostly aimed at DNN watermarks (passive copyright verification). In response to the new direction discussed in this study --- active authorization control, what attacks and bypass mechanisms may exist?
\item How to evaluate DNN copyright protection methods based on the needs of commercial applications, in particular, how to evaluate the new direction --- DNN active authorization control?
\end{itemize}

In order to achieve the aforementioned objectives, 
it becomes imperative to study and solve the following technical difficulties/challenges in active authorization control schemes:
(i) how to construct unique users' identities; (ii) how to generate imperceptible users' fingerprints to resist attacks or fingerprint leakage; (iii) how to enable the DNN model to distinguish between authorized and unauthorized users; (iv) how to distinguish different authorized users; (v) how to control the functionality and performance of DNN models differently based on different users.

\begin{figure*}[!htbp]
\centering
\includegraphics[width=0.85\textwidth]{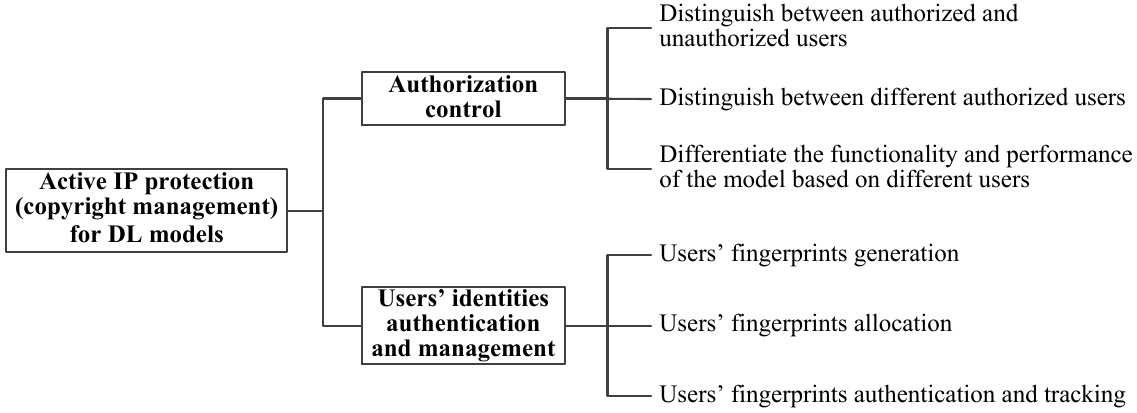}
\caption{The framework of active IP protection for DL models.}
\label{fig_Framework}
\end{figure*}

An active DL model copyright protection method
possesses the capacity to proactively thwart piracy.
In addition, most existing methods embed watermarks or extract fingerprints from the model to verify the ownership of the model, but do not consider copyright management.
Addressing the pressing requirement for robust commercial DNN copyright protection necessitates the effective authentication and management of user identities.
To this end, the DL model active IP protection (also known as copyright management) framework includes two modules: user identity authentication and management, and active authorization control, as shown in Fig.~\ref{fig_Framework}.

The \textbf{active authorization control} module involves:
\begin{itemize}
\item  Distinguish between authorized and unauthorized users: The model must differentiate between  authorized and unauthorized users.
\item Distinguish between different authorized users: The model should be able to distinguish between different authorized users.
\item  Differentiate the functionality and performance of the DL model based on different users: The DL model should be able to provide different functions and performance based on the identity of users. Authorized users can use the DL model normally and achieve normal performance, while unauthorized users cannot use the model or can only achieve extremely low model performance.
\end{itemize}

The \textbf{user identity authentication and management} module involves:
\begin{itemize}
\item User fingerprint generation: Generate a unique user fingerprint/identity for each authorized user. Besides, in order to resist attacks or fingerprint leakages, the ideal user fingerprint should be visually imperceptible.
\item User fingerprint allocation: Each authorized user is assigned a unique user fingerprint, which should be able to resist collusion attacks, forgery attacks, etc.
\item User fingerprint authentication and tracking: The model can extract the user's fingerprint from the input for authentication, and can track the user's identity.
\end{itemize}

To address the pressing need for commercial DNN copyright protection, and to overcome current research limitations, it is imperative to explore proactive DNN copyright protection mechanisms. These mechanisms should be based on active authorization control and user identity authentication and management (copyright management). Key components of this endeavor include:
(i) Study unique user fingerprint generation, allocation, authentication and management methods for DNN; (ii) Establish an active authorization control mechanism that distinguishes between authorized and unauthorized users for DNN models; (iii) Study the attack and bypass mechanisms against DNN active authorization control (all existing attacks are targeted at DNN watermark), and carry out hierarchical attack mechanism research according to the capabilities of different attackers; (iv) Establish a comprehensive evaluation framework for DNN copyright protection methods, covering basic functional metrics and attack resistance metrics, with a particular focus on evaluation methods for newly proposed active authorization control mechanisms (the evaluation metrics for active authorization control are different from those of existing DNN watermarks). Active copyright protection can solve bottlenecks such as passivity in DNN watermarks and difficulty in copy management. It also addresses intricate issues such as active authorization control, authentication and tracking of DNN user identities, thus can lay a theoretical foundation for constructing an active, secure, and robust DNN copyright protection mechanism, and provide theoretical and technical support for DNN's commercial applications.

\subsection{Unique Metrics for Active DL IP Protection}
The assessment of existing DNN IP protection methods has predominantly revolved around metrics tailored to DNN watermarking, leaving a research gap in terms of evaluation metrics for DNN active authorization control. This paper introduces a novel aspect by formulating evaluation metrics for DNN active authorization control, distinct from conventional DNN watermarking approaches. Furthermore, the existing focus on functional metrics has often overshadowed the need for robust attack-resistance metrics.
This paper discusses the unique metrics for active DL IP protection, which are divided into basic functional metrics and attack-resistance metrics. The proposed functional metrics for active authorization control include cost, user authentication success rate, model performance that authorized users can obtain, model performance that unauthorized users can obtain, uniqueness of user identity, number of supported users, convenience of deployment, knowledge of the target model, interpretability, etc.
The proposed attack-resistance metrics for active authorization control include user identity concealment, resistance to sample modification attacks, user identity unforgeability, resistance to adaptive attacks, and resistance to reverse analysis, etc. The unique metrics for active DL IP protection proposed in this study can provide evaluation references for subsequent related work in this field, and can also provide feedback on improving active copyright protection mechanisms based on the evaluation results.

Specifically, this paper proposes the following unique metrics for DNN active authorization control, which are different from the common metrics for DNN watermarks.

\textbf{Basic functional metrics} include:
\begin{itemize}
\item  \textit{Overhead}: The overhead introduced by the active authorization control scheme should be low or no overhead, within the range that model owners and users can afford.
\item \textit{Success rate of user identity authentication}:
      In the active authorization control scheme, user identity authentication and management are introduced, and the success rate of user identity authentication needs to be evaluated.
\item \textit{Model performance that authorized users can obtain}:
      In the active authorization control scheme, the model usage and model performance of authorized and unauthorized users are differentiated, and authorized users should be able to achieve high model performance.
\item \textit{Model performance that unauthorized users can obtain}:
      In the active authorization control scheme, the model usage and performance of authorized and unauthorized users are differentiated, and unauthorized users should achieve extremely low model performance, or even be unable to use the model.
\item \textit{Uniqueness of user identity}:
      In the active authorization control scheme, user identity authentication and management are included, and each user's identity should be unique.
\item \textit{Number of supported users}:
      In the active authorization control scheme, unique identities are generated for users for authorization control, and the scheme should be able to support a large number of users' identities generation.
\item \textit{Convenience of deployment}:
      Does the active authorization control scheme require training from scratch to be implemented, or can it be implemented through fine-tuning?
\item \textit{Knowledge of the target model}:
      The ideal scenario is that the solution can be applied to black-box scenarios, rather than only white-box scenarios.
\item \textit{Interpretability}:
      The interpretability of the protection scheme is a requirement in some commercial applications.
\end{itemize}

\textbf{Attack-resistance metrics} include:
\begin{itemize}
\item \textit{The concealment of user identity}:
      In active authorization control schemes, in order to prevent attacks or leakage of user identities, the user identities embedded in images and samples should be visually imperceptible and have good concealment.
\item \textit{Resistance to sample modification attacks}:
      In active authorization control schemes, legitimate samples or samples containing user identities should be able to resist malicious modifications to the samples.
\item \textit{User identity unforgeability}:
      In active authorization control schemes, the identity of authorized users should be unforgeable, making it difficult for attackers to forge a legitimate user identity, and the user identity forged by the attacker cannot pass the authentication of the model.
\item \textit{Resistance to adaptive attacks}:
      In the worst-case scenario, the attacker knows the mechanism of active copyright protection, and the active authorization control scheme should be able to resist such adaptive attacks.
\item \textit{Reverse analysis resistance}:
      The attacker attempts to reverse the active authorization control mechanism and disrupt it, and the active IP protection scheme should be able to resist reverse analysis.
\end{itemize}

\section{Overview of Existing DL Active Copyright Protection Work} \label{sec_review}

The majority of existing DL copyright protection methods fall within the category of passive DNN watermark verification. These techniques are reactive in nature, as they verify model copyright after piracy or infringement have occurred. However, there is a growing body of research dedicated to proactive DL copyright protection, aiming to prevent piracy and infringement proactively.
These proactive DL active copyright protection efforts mostly focus on providing active authorization control for the model, which can distinguish between authorized and unauthorized users (authorized users can use the model or achieve high performance, while unauthorized users cannot use the model or achieve low performance), but do not consider user identity authentication and management, which cannot track the identity of authorized users and distinguish different authorized users. In the following, we will review the existing DL active copyright protection work from two aspects: (i) active authorization control; (ii) both active authorization control and user identity management. The overview of existing active DL copyright protection work is shown in Table \ref{tab1}.

\begin{table*}[htbp]
  \centering
  \caption{Existing Active DL Copyright Protection Work}
    \begin{tabular}{|p{7em}<{\centering}|p{23em}<{\centering}|p{7em}<{\centering}|p{7em}<{\centering}|}
    \hline
         \multirow{3}{*}{Work} & \multirow{3}{*}{Mechanism} & Authorization control & User identity authentication and management \\
    \hline
            \multirow{2}{*}{Chen and Wu \cite{chen2018protect}} & preprocess the input via a conversion module based on adversarial perturbation & \multirow{2}{*}{\Checkmark}     & \multirow{2}{*}{--} \\
    \hline
    \multirow{2}{*}{Fan et al. \cite{fan2021deepipr}} & unless a valid passport is provided, the DNN model will not function properly & \multirow{2}{*}{\Checkmark}     & \multirow{2}{*}{--} \\
    \hline
    \multirow{2}{*}{Ren et al. \cite{ren2022protecting}} & lock the model, and only specific tokens can unlock the model & \multirow{2}{*}{\Checkmark}    & \multirow{2}{*}{--} \\
   \hline
    Lin et al. \cite{lin2020chaotic} & model encryption based on chaotic weights & \Checkmark     & -- \\
    \hline
    AprilPyone and Kiya \cite{AprilPyoneK21} & applying block transformation with a key to feature maps for authorization control & \multirow{2}{*}{\Checkmark}    & \multirow{2}{*}{--} \\
    \hline
    \multirow{3}{*}{Xue et al. \cite{xue2022advparams}} & based on gradients, the few parameters that have the greatest impact on model performance are slightly perturbed, thereby encrypting the model & \multirow{3}{*}{\Checkmark}     & \multirow{3}{*}{--} \\
    \hline
    \multirow{2}{*}{Luo et al. \cite{Hierarchical}} & perturb the output of the model to varying degrees to achieve hierarchical performance & \multirow{2}{*}{\Checkmark}     & \multirow{2}{*}{--} \\
    \hline
    \multirow{3}{*}{Pan et al. \cite{pan2022device}} & the model weights were encrypted/decrypted based on permutation and diffusion, and a key bound to the device was generated based on PUF & \multirow{3}{*}{\Checkmark}     & \multirow{3}{*}{--} \\
    \hline
    Chakraborty et al. \cite{chakraborty2020hardware} & confusion framework for hardware NN, which requires authorization through the key in the hardware & \multirow{2}{*}{\Checkmark}     & \multirow{2}{*}{--} \\
    \hline
    \multirow{2}{*}{Xue et al. \cite{xue2020active}} & a user fingerprint management and DNN authorization control framework based on multi-trigger backdoor & \multirow{2}{*}{\Checkmark}     & \multirow{2}{*}{\Checkmark} \\
    \hline
    \multirow{3}{*}{Tang et al. \cite{tang2020DSN}} & teacher-student model, in which the customer DNN can only function properly when the customer enters a valid serial number & \multirow{3}{*}{\Checkmark}     & \multirow{3}{*}{\Checkmark} \\
    \hline
    \multirow{2}{*}{Wang et al. \cite{WangXZC22}} & embed different backdoors into the model to generate a number of user model instances for different users & \multirow{2}{*}{\Checkmark}    & \multirow{2}{*}{\Checkmark} \\
    \hline
    \multirow{2}{*}{Chen et al. \cite{chen2019deepmarks}} & encoding each constructed user fingerprint in the probability density function of the weights & \multirow{2}{*}{\Checkmark}     & \multirow{2}{*}{\Checkmark} \\
    \hline
    \multirow{3}{*}{Xue et al. \cite{xue2023activeguard}} & using carefully crafted adversarial examples (with specific categories and specific confidences) as user fingerprints & \multirow{3}{*}{\Checkmark}    & \multirow{3}{*}{\Checkmark} \\
    \hline
    \multirow{3}{*}{Xue et al. \cite{XueSunAPIN}} & using an additional class and user identity information embedded through steganography to support both user identity authentication and model ownership verification & \multirow{3}{*}{\Checkmark}     & \multirow{3}{*}{\Checkmark} \\
    \hline
    \multirow{2}{*}{Fan et al. \cite{PCPT}} & using an additional class, authorization control, and image perceptual hash & \multirow{2}{*}{\Checkmark}     & \multirow{2}{*}{\Checkmark} \\
    \hline
    \multirow{3}{*}{Wu et al. \cite{aicas}} & reverse and multiple use of sample-specific backdoors for authorization control and embed imperceptible user identity information & \multirow{3}{*}{\Checkmark}    & \multirow{3}{*}{\Checkmark} \\
    \hline
    \end{tabular}%
  \label{tab1}%
\end{table*}%

\subsection{Active Authorization Control}
Chen et al. \cite{chen2018protect} propose a DNN access control framework so that only authorized users can use the model normally. They design a conversion module based on adversarial examples to provide authorized input, where adversarial disturbance is added to the input. Authorized users can use conversion modules to preprocess inputs, resulting in high model performance. In contrast, unauthorized users who directly provide input to the model will result in poor model performance.

Fan et al. \cite{fan2021deepipr} propose embedding specific passports into neural networks. The inference performance of the DNN model is adjusted based on the provided passport, that is, the inference performance of the pre-trained DNN model will remain unchanged in the presence of a valid passport, or significantly deteriorate due to passport modification or forgery. Unless a valid passport is provided, the DNN model will not function properly, thus preventing illegal use of the model.

Ren et al. \cite{ren2022protecting} propose a model-locking scheme for deep learning, aimed at preventing attackers from achieving high prediction accuracy even if they pirate the model. If a specific token does not exist in the input, the locked DNN model will provide poor prediction accuracy. When the input contains the specifically designed authorization token, the model can make normal predictions.

Lin et al. \cite{lin2020chaotic} propose a chaotic weight framework based on chaotic mapping theory, which achieves an encryption effect by exchanging weight positions, making the kernel of convolutional or fully connected layers chaotic. Unless the model is decrypted, an incorrect prediction result will be returned. These encryption-based implementations may affect the performance of the model or introduce high overhead.

AprilPyone and Kiya \cite{AprilPyoneK21} propose a protection method with keys for convolutional neural network (CNN) models, which applies block transformations with keys to feature maps, enabling authorized users to achieve high classification accuracy while unauthorized users achieve low classification accuracy.

\begin{figure}[!htbp]
\centering
\includegraphics[width=0.5\textwidth]{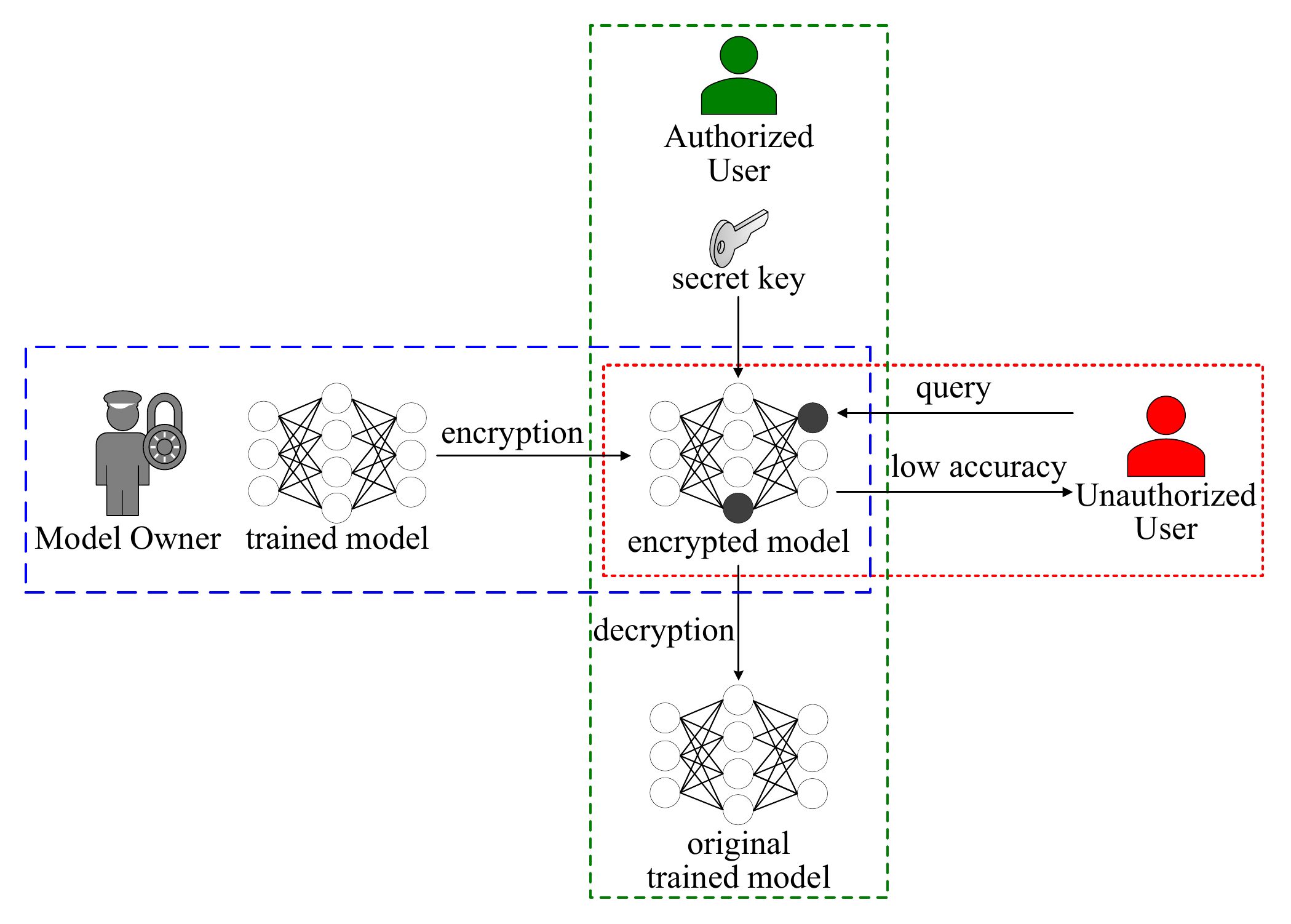}
\caption{DNN active authorization control based on parameter perturbation \cite{xue2022advparams}.}
\label{fig_AdvParam}
\end{figure}

Xue et al. \cite{xue2022advparams} propose an active DNN copyright protection based on parameter perturbation, as shown in Fig. \ref{fig_AdvParam}. The extremely small number of parameters that have the greatest impact on model performance are slightly perturbed based on gradient. By encrypting a very low number of parameters, the accuracy of the model can be significantly reduced. Authorized users can decrypt models in MLaaS and achieve high-accuracy model performance.

Luo et al. \cite{Hierarchical} propose a multi-user hierarchical authorization for CNN, which can help owners control output results based on different levels of access permissions. They refer to differential privacy and use the Laplace mechanism to perturb the output of the model to different degrees to achieve the hierarchical performance.

Pan et al. \cite{pan2022device} encrypt/decrypt the model weights based on permutation and diffusion to achieve IP protection, and a key bound to the device was generated based on Physical Unclonable Function (PUF). This method is targeted at DL hardware and has a high overhead.

Chakraborty et al. \cite{chakraborty2020hardware} propose a hardware neural network confusion framework  that requires a key in hardware for authorization to be used. The model owner first uses the key-based backpropagation to train the DNN architecture, which blurs the weight space of the model, and then hosts the deep learning model in the shared platform. Only authorized users with hardware-trusted roots (on-chip memory with an embedded key) can use the deep learning model. The above-mentioned hardware DNN copyright protection work focuses on the copyright protection of DNN hardware devices, and requires hardware platforms for support, such as hardware trusted roots, resulting in high costs.

\subsection{User Identity Authentication and Management}

The DL copyright protection methods discussed earlier prioritize authorization control but do not address user identity authentication and management. Consequently, they do not fully align with the requirements of commercial DL copyright management. Furthermore, they are also susceptible to attacks initiated by dishonest users, such as collusion attacks. Recently,
a limited body of research focus on the development of active copyright protection techniques that support both authorization control and user identity authentication and management.

Xue et al. \cite{xue2020active} propose a user fingerprint management and DNN authorization control framework based on multi-trigger backdoors \cite{xue2020one}. First, $N$ sub-backdoors are implanted in the DNN model. Then, a small group of images with $n$ ($n<N$) subbackdoor triggers corresponding to moderate confidence are assigned to authorized users as their unique fingerprints. Only the model owner has all the $N$ backdoor triggers corresponding to high confidence, which are used to verify the ownership of the model. In this way, both user identity authentication and ownership verification can be achieved.

Tang et al. \cite{tang2020DSN} propose a watermarking method based on deep serial numbers. First, a private teacher DNN model was trained, and then their knowledge was extracted and transferred to a series of customized student DNN models. During the distillation process, each customer DNN has a unique serial number, which is an encrypted 0/1 trigger pattern. The customer DNN can only function properly when the customer enters a valid serial number.

Wang et al. \cite{WangXZC22} propose a buyer-traceable DNN model IP protection method. They embed different backdoors defined by owners into the model through training data poisoning with dirty labels, and generate a large number of user model instances that maintain accuracy for different buyers. Each backdoor training model can be triggered by a set of special validation images. Model instances sold to different buyers can be uniquely distinguished by triggering their specific backdoors.

Chen et al. \cite{chen2019deepmarks} state that in large-scale model distribution systems, multiple users can use their respective watermarked models to collaborate and construct an unmarked high-performance model, known as fingerprint collusion attacks. They propose a secure fingerprint framework for digital rights management for deep learning models. They design unique fingerprints for individual users. By combining the specific regularization loss of fingerprints during DNN retraining, each constructed fingerprint is encoded in the probability density function of the weight, so that the owner can verify the information of the model owner and the unique identity of the user, and can resist fingerprint collusion attacks.

Xue et al. \cite{xue2023activeguard} propose an active DNN copyright protection method through adversarial example-based user fingerprinting. It can provide DNN with active authorization control, user identity management, and ownership verification. Specifically, they use carefully crafted adversarial examples (specific categories with specific confidence levels) as users' fingerprints. Authorized users can input their fingerprints into DNN for authentication and obtain normal usage, while unauthorized users will achieve significantly poor model performance. In addition, model owners embed watermarks into the weights of DNN for ownership verification.

\begin{figure*}[!htbp]
\centering
\includegraphics[width=0.88\textwidth]{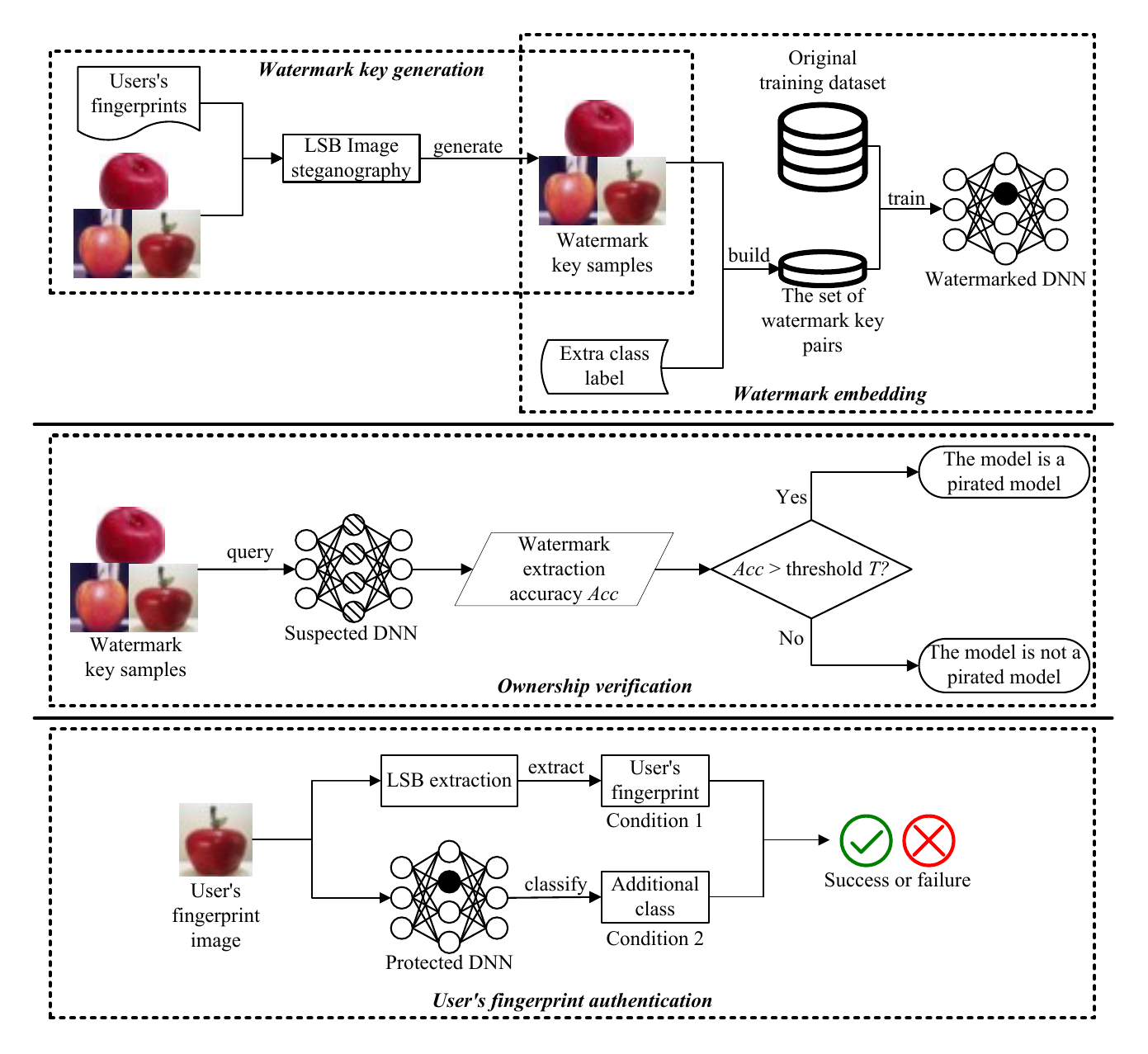}
\caption{DNN active authorization control and user identity authentication based on extra class and steganography \cite{XueSunAPIN}.}
\label{fig_APIN}
\end{figure*}

Xue et al. \cite{XueSunAPIN} propose a DNN model active copyright protection method based on hidden backdoor and user identity authentication, as shown in Fig. \ref{fig_APIN}. They utilize an additional class to support both user identity authentication and model ownership verification. In order to embed hidden backdoor (watermark) in DNN, a small number of images are selected from outside the training set as watermark key samples. Then, the fingerprint information representing the user's identity is embedded into the watermark key sample, and an additional class label is assigned to the watermark key sample \cite{XueSunAPIN}.
Next, they add the watermark key sample to the training dataset of DNN for model training. After training, DNN will be embedded with the watermark. In the testing phase, DNN always predicts watermark key samples as the additional class. In addition, since the watermark key samples originate from samples outside the dataset (rather than samples with obvious watermark patterns), and the watermark is embedded through an additional class, a strong connection is established between the watermark key samples and the watermark, which can resist query modification attack. In order to support user authentication, steganography is used to hide user fingerprint information into the watermark key sample, so that the user fingerprint information is imperceptible \cite{XueSunAPIN}. In the user authentication stage, the watermark key sample is used as the user fingerprint and submitted to DNN for identity verification.

Fan et al. \cite{PCPT} propose an active copyright protection and traceability framework for DNN. Combining authorization control strategies and image perceptual hash algorithms, an authorization control center is constructed. In this framework, the detector detects whether the key image input by the user is legal, and the authenticator verifies whether the user's identity information is legal.

\renewcommand{\dblfloatpagefraction}{0.8}

\begin{figure*}[!htbp]
\centering
\includegraphics[width=0.98\textwidth]{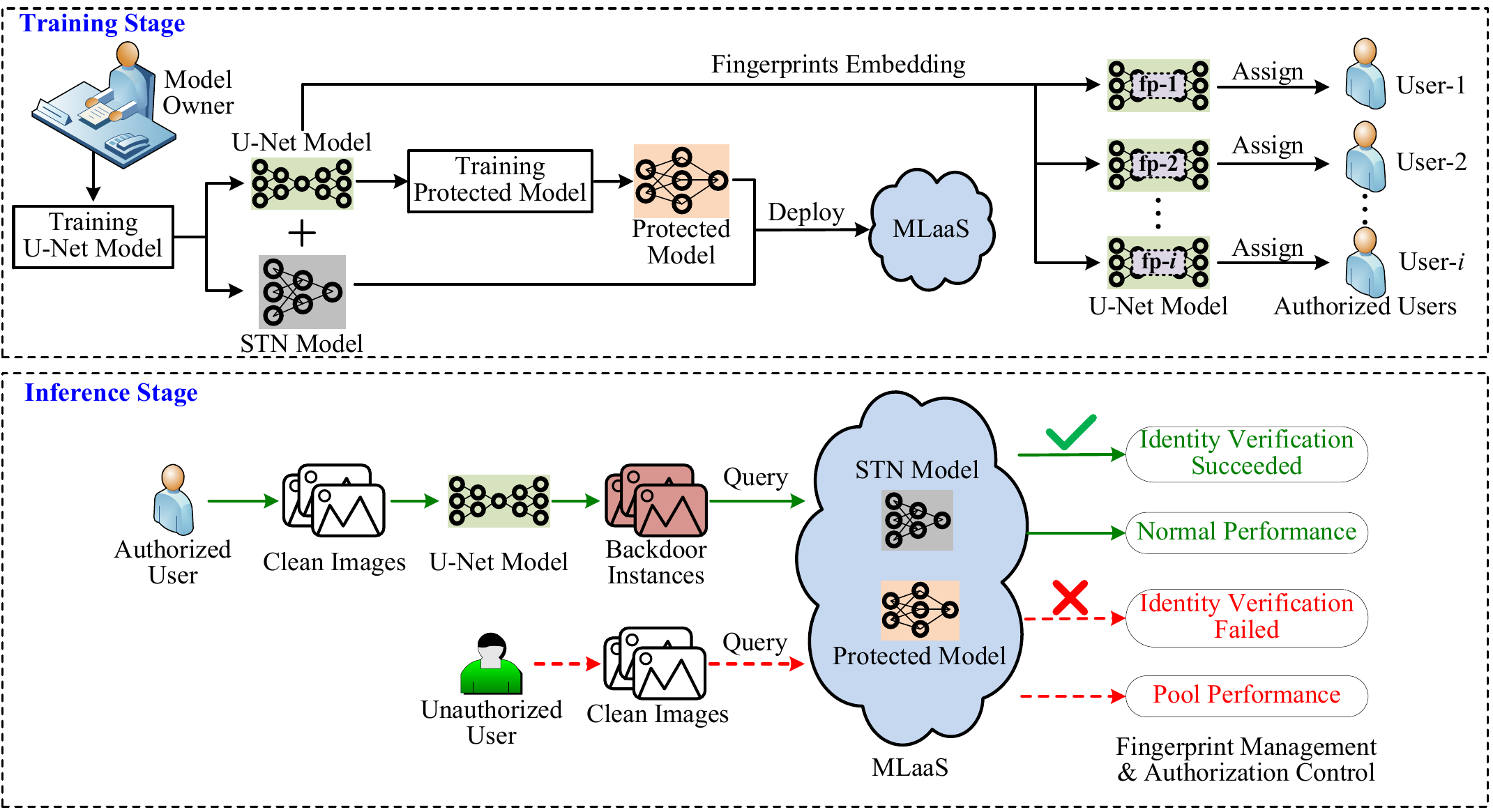}
\caption{Sample-specific backdoor based active DNN IP protection \cite{aicas}.}
\label{fig_TOMM}
\end{figure*}

Wu et al. \cite{aicas} use a specific backdoor to provide both authorization control and user identity management for the DNN model, as shown in Fig. \ref{fig_TOMM}. The backdoor trigger is used as an authorization key in this scheme, and there are hidden fingerprints that can be extracted to verify the user's identity. They reversely utilize the characteristics of sample-specific backdoor attacks, where backdoor triggers are used to guide images from the wrong class to the correct class, thereby achieving authorization control \cite{aicas}. In this way, the model copyright protection method can achieve normal model performance only for authorized users (who have the U-Net model), and unauthorized users cannot use the model normally. At the same time, the backdoor trigger is not a fixed pattern, but is generated through the U-Net model. Therefore, the user with the trigger generation model can use the target model normally. In this method, for different clean images, the corresponding trigger is imperceptible and image related, that is, the generated backdoor trigger is related to the content of the corresponding clean image, and different images correspond to different backdoor triggers. In addition, each backdoor trigger hides a unique fingerprint corresponding to an authorized user, and the model owner can verify and track the user's identity by extracting the fingerprint from the backdoor instance. When distributing the U-Net model to authorized users, the U-Net model can be set to embed any string into the image. They hide the user's unique fingerprint as a string into the U-Net model and assign this U-Net to the corresponding authorized user \cite{aicas}. In the inference stage, when authorized users generate backdoor instances through the U-Net model, their fingerprints are embedded into the image along with the backdoor trigger. Subsequently, the model owner can extract this fingerprint from the submitted backdoor instance to verify and track the user's identity.

\subsection{Challenges}
As reflected by the discussions above, in the realm of DNN copyright protection, there are several pressing challenges and bottleneck difficulties that require attention:
\begin{itemize}
\item Most existing DNN copyright protection methods rely on passive verification, which means they ascertain the ownership of DNN models only after instances of piracy and infringement have occurred.
This post-verification method cannot proactively prevent the occurrence of piracy and infringement. To achieve large-scale commercial applications of the DNN model, there is an urgent need for active copyright protection mechanisms.
\item Most existing DNN copyright protection methods only focus on verifying model copyright, without managing the unique identity of users, thereby lacking effective copyright management, which is a necessary and core requirement for commercial DNN copyright protection.
    Further, they are vulnerable to several attacks, such as collusion attacks.
\item The various attacks considered in existing work are all aimed at DNN watermarks (post-verification copyright protection). Regarding the new direction studied in this paper --- active authorization control, the possible attack and bypass mechanisms are still blank.
\item Lack of a systematic evaluation methodology. As an emerging cutting-edge research direction, DNN copyright protection is still in its infancy and lacks a comprehensive evaluation method. This gap is particularly evident in the case of active authorization control, where evaluation metrics and methods are yet to be developed.
\end{itemize}

\section{Attacks Faced by Active DL IP Protection} \label{sec_attacks}

The existing attacks on DNN copyright protection are all targeted at DNN watermarks, including model modification attacks, evasion/removal attacks, and some more aggressive attack scenarios \cite{tai}.

This paper focuses on studying attacks and bypass mechanisms against DL active authorization control, including but not limited to the following:

\textbf{Query modification attack} \cite{NambaS19}: Query modification attack is a strong attack method. This attack aims to use an autoencoder to detect and remove watermarks in the watermark key sample, thereby invalidating the watermark query during ownership verification. We believe that autoencoders can be used to attack watermark key samples. The work \cite{XueSunAPIN} claims that their scheme is robust to query modification attacks for the following reasons. First,
in the scheme, the watermark key sample is generated from clean images in a different dataset. This design choice ensures that the watermark trigger pattern becomes virtually indistinguishable from that of clean images. As a result, attackers are unable to detect watermarks by searching for distinctive watermark trigger patterns within the images.
Second,
the scheme establishes a robust connection between the watermark key sample and an additional class label. This connection is resilient, so even if the watermark key sample undergoes slight modifications as a result of the autoencoder attack, the DNN can still classify the altered watermark sample into the additional class \cite{XueSunAPIN}.

\textbf{Adaptive attack}: Adaptive attack refers to the fact that an attacker knows the details of the protection mechanism and attempts to bypass it, which is the worst-case attack scenario. We will take the active authorization control mechanism in the work \cite{aicas} as an example for discussion. In practical applications, unauthorized users are unable to access protected models and U-Net models. In order to discuss the robustness of the scheme, they consider the worst-case scenario, assuming that the attacker knows that the proposed method is based on backdoor and unintentionally obtains a pair of clean images and corresponding backdoor instances. In this way, the attacker attempts to obtain the corresponding trigger by subtracting the clean image from the backdoor instance. Then the attacker adds the trigger to another clean image to forge a ``backdoor instance''. The analysis of the work \cite{aicas} shows that the clean accuracy of the protected model under adaptive attacks remains at a relatively low level. Therefore, these ``fake backdoor instances'' will still be predicted as incorrect classes by the protected model. The reason is that this scheme generates different triggers for different clean images. Meanwhile, the generated sample-specific trigger is only effective for the corresponding clean image. Therefore, when a trigger generated for a clean image $x_1$ is used for another clean image $x_2$, the trigger will be invalid, and the protected model will still consider the image as a clean image and output the wrong class. In summary, when attackers use forged triggers to perform adaptive attacks, the method \cite{aicas} still has robustness.

\textbf{Compression attack}: Attackers may use image compression to disrupt validation or authentication samples. Using the scheme in the work \cite{aicas} as an example, the image processed by the U-Net model in the scheme may be compressed during transmission, and the compression operation will remove redundant pixels in an image. The work \cite{aicas} uses common JPEG compression method to evaluate the robustness of the scheme in the face of compression attacks. The image processed by U-Net is subjected to JPEG compression, and then the compressed backdoor instance is fed into the protected model for prediction. It is shown that the model prediction accuracy obtained by authorized users under JPEG compression attacks is close to the accuracy obtained without JPEG compression attacks. During the process of image transmission, even if the image is compressed by JPEG, authorized users can still achieve normal prediction accuracy. In addition, even if the image has been compressed using JPEG compression, the method can still effectively extract and verify the user's fingerprint. Therefore, the method \cite{aicas} is robust to JPEG compression attacks.

\textbf{Collusion attack}: A group of authorized users with the same host DNN and different fingerprints may engage in collusion attacks, attempting to construct a model with the same functionality \cite{chen2019deepmarks}, or colluding with unauthorized users to provide authorized use for multiple unauthorized users.

\textbf{User fingerprint forgery}: Attackers attempt to forge legitimate user fingerprints in the input image to pass  identity authentication and obtain model authorization. Taking the work \cite{XueSunAPIN} as an example, if an attacker wants to successfully forge a fingerprint, he needs to know the additional class, the steganography used, and the specific format of the fingerprint. Due to the fact that the above information is also black-boxed and unknown to authorized users, it is difficult for attackers to obtain the above information, making it difficult to forge legitimate user fingerprints. Taking the work \cite{aicas} as an example, an attacker wants to successfully forge a user fingerprint. However, the user fingerprint is associated with each sample, and is visually imperceptible. Authorized users cannot obtain fingerprint information, and if they are lucky enough to crack the corresponding fingerprint of one sample, it cannot be applied to other samples. Therefore, it is difficult for attackers to forge legitimate user fingerprints.

\textbf{Cracking authorization control module}: Attackers attempt to crack the core of an active DL IP protection scheme --- the active authorization control module. Taking the work \cite{aicas} as an example, attackers attempt to crack the specially trained U-Net module. However, the U-Net module is jointly trained with the model, and it is specially trained by four loss functions. Moreover, a unique fingerprint is embedded in the U-Net module. Cracking and forging such a U-Net model is more difficult than stealing a commercial DNN model.

From a defense perspective, strategies for enhancing the robustness of these DL IP protection schemes include (but are not limited to): (i) Customized defense strategies for specific attacks. For example, to combat watermark removal attacks, it is suggested to embed watermarks by training from scratch, and embedding the watermarks as inseparable from the main task, so that if the attacker wants to remove the watermark, it will inevitably damage the performance of the original model; (ii) A defense mechanism based on generative adversarial networks, in which the training process of embedding watermarks is the generator, while the detector that integrates multiple watermark detection methods is the discriminator, which can embed hidden watermarks with strong attack resistance. Overall, what attacks might exist  and how to defend against them remain open questions.

\section{Future Directions}
\label{sec_directions}

\subsection{An Active Authorization Control Method against Query Modification Attack with Lossless Model Performance}
In existing work, embedding watermarks in the model may affect the performance of the model on the main task. In the active authorization control scheme, how to ensure that the constructed user identity authentication-based active authorization control method does not affect the performance of the model on the main task, and how to construct key samples or authentication samples that can resist query modification attack \cite{NambaS19}, are key challenges.

\subsection{Distinguish between Different Authorized Users}
Most of the existing DL model active copyright protection schemes focus on providing active authorization control for model owners, which can distinguish between authorized and unauthorized users, but do not consider user identity authentication and management, which cannot track the identity of authorized users and distinguish different authorized users.

\subsection{Unique and Imperceptible Fingerprint Generation Method for Authorized Users, and Model Performance Differentiation Control Method}
Most of the existing DNN copyright protection methods are passive which verify copyright after piracy and infringement occur. The active DNN copyright protection method can lock the model and actively prevent piracy. Besides, most existing methods embed watermarks or extract fingerprints from the model to verify the ownership of the model, but do not consider copyright management. In order to meet the urgent need for commercial DNN copyright protection, it is necessary to effectively authenticate and manage the identities of users.
In the active authorization control scheme, how to construct unique and imperceptible fingerprint generation method for authorized users, and how to control the functionality and performance of DNN models differently based on different users, are the difficulties/challenges to be solved in the future.

\subsection{Attack Mechanism for Active Authorization Control}
Most of the existing DNN copyright protection work has evaluated the robustness of model modification, while a small amount of work has considered watermark removal attacks or evasion attacks. Regarding the new direction of this study --- active authorization control, the research on possible attacks and bypass mechanisms is still lacking. What targeted attack methods will attackers adopt under the DNN active authorization control framework? How active authorization control schemes can overcome strong attack methods such as adaptive attacks, query modification attacks, compression attacks, collusion attacks, user fingerprint forgery attacks, and authorization control module cracking attacks, is a future research direction.

\subsection{Evaluation Methods for Active Authorization Control Considering Attack Resistance}
There is still a lack of research on the evaluation of DNN active authorization control. In addition, most of the existing evaluations of DNN copyright protection work focus on the functional metrics of DNN watermarks, and there is insufficient research on attack-resistance metrics. For active authorization control frameworks, how to construct evaluation methods and metrics that consider attack resistance is a future research direction.

\subsection{Active Authorization Control for Datasets}
In addition to active authorization control for the DL models, there have been very few researchers recently paying attention to the authorization control of the datasets \cite{XueDataset, chaixiuli}. Datasets also have high value and are widely used in more and more scenarios. How to construct authorization control for the usage of datasets is also a potential research direction.

\section{Conclusion}
\label{sec_conclusion}
For the first time, this review focuses on cutting-edge active DL copyright protection methods, analyzes current bottlenecks, discusses the goal and connotation of proactive copyright protection --- providing active copyright protection (actively prevent piracy) and copyright management (user identity authentication) functions, and calls for studies on the attack mechanisms and evaluation of active DL authorization control. Studying proactive DNN copyright protection is expected to break through the bottlenecks and cutting-edge challenges faced by deep neural network copyright protection, and provide effective copyright protection mechanisms and lay a theoretical foundation for the commercial implementation of deep neural network models, which has significant significance.

\bibliographystyle{IEEEtran}
\bibliography{ref}

\end{document}